# Structural, Elastic, and Electronic Properties of Recently Discovered Ternary Silicide Superconductor Li$_2$IrSi$_3$: An *ab-initio* Study

M. A. Hadi[1], M. A. Alam[2], M. Roknuzzaman[3], M. T. Nasir[1], A. K. M. A. Islam[1,4], and S. H. Naqib[1*]

[1]Department of Physics, University of Rajshahi, Rajshahi-6205, Bangladesh
[2]Department of Physics, Rajshahi University of Engineering and Technology, Rajshahi-6204, Bangladesh
[3]Department of Physics, Jessore University of Science and Technology, Jessore-7408, Bangladesh
[4]International Islamic University Chittagong, 154/A College road, Chittagong, Bangladesh

**Abstract**

The structural, elastic, and electronic properties of the very recently discovered ternary silicide superconductor, Li$_2$IrSi$_3$, have been calculated using *ab-initio* technique. We have carried out the plane-wave pseudopotential approach within the framework of the first-principles density functional theory (DFT) implemented within the CASTEP code. The calculated structural parameters show a reasonable agreement with the experimental results. Elastic moduli of this interesting material have been calculated for the first time. The electronic band structure and electronic energy density of states indicate the strong covalent Ir-Si and Si-Si bonding which lead to the formation of the rigid structure of Li$_2$IrSi$_3$. Strong covalency give rise to a high Debye temperature in this system. We have discussed the theoretical results in detail in this paper.

**Keywords:** Silicide superconductor; Crystal ctructure; Elastic properties; Electronic structures

## 1. Introduction

Transition metals often form covalent bonds with tetravalent silicon, because their electronegativity is comparable to that of silicon. The covalent character plays a vital role in realizing phonon-mediated superconductors with a high $T_c$. Transition metal silicides have been already identified as potential superconductors. For instance, V$_3$Si exhibits superconductivity at $T_c$ = 17 K [1]. In addition, the tunability of chemical potential by changing the transition metals may provide an additional channel in designing the ideal electronic structure [2] leading to higher $T_c$. For these reasons, the searching of new superconductors in doped covalent metals has received substantial attention. Very recently, using the conventional solid state reaction procedure, a new ternary transition metal silicide Li$_2$IrSi$_3$ was synthesized by Hirai et al. [2]. This compound crystallizes in the hexagonal system with space group P6$_3$/mmc (No. 194) and shows weak-coupling phonon-mediated superconductivity below 3.7 K. The crystal structure of Li$_2$IrSi$_3$ is composed of Si triangles connected by Ir atoms, resulting in a three-dimensional network of covalent bonds. The antiprisms, IrSi$_6$ stacking along the c-axis are connected by Si triangles.

At the same time, Pyon et al. [3] prepared the polycrystalline samples of Li$_2$IrSi$_3$ by the arc-melting method and identified as a noncentrosymmetric trigonal structure with the space group *P*31*c* (No. 159). Resistivity and magnetization measurements reveal that Li$_2$IrSi$_3$ is a type-II superconductor with a $T_c$ = 3.8 K. The crystal structure of Li$_2$IrSi$_3$ was found to consist of a planar kagome network of silicon atoms with Li and Ir spaced at unequal distances between the kagome layers, giving rise to a polar structure along the c-axis.

In this paper, we aim to explore the elastic and electronic properties of the newly discovered transition metal silicide superconductor Li$_2$IrSi$_3$ by means of first-principles calculations. No study of the elastic and mechanical properties of Li$_2$IrSi$_3$ has been done yet. The outline of the paper is as follows: In section 2, a brief description of computational formalism used in this study has been presented. All the results obtained in the present investigation are given in section 3, while section 4 consists of discussion and concluding remarks.

---

[*] Corresponding author: salehnaqib@yahoo.com

## 2. Computational Procedures

The first-principles calculations reported herein were performed using the Cambridge Serial Total Energy Package (CASTEP) code [4], which is based on the density functional theory (DFT) [5, 6] within the plane-wave pseudopotential approach. The Kohn-sham equations were solved using the Perdew-Burke-Ernzorhof generalized gradient approximation (PBE-GGA) [7] for the exchange-correlation energy. Vanderbilt-type ultrasoft pseudopotentials [8] were used to model the electron-ion interactions. Throughout the calculations, a plane-wave cutoff energy of 500 eV was chosen to determine the number of plane waves in expansion. The crystal structures were fully relaxed by the Broyden-Fletcher-Goldfrab-Shanno (BFGS) minimization technique [9]. Special $k$-points sampling integration over the Brillouin zone was employed by using the Monkhorst–Pack scheme [10] with the 15×15×8 mesh. This set of parameters uses the tolerance in the self-consistent field calculation of $5\times10^{-7}$ eV/atom, the change in total energy of $5\times10^{-6}$ eV/atom, the maximum force of 0.01 eV/Å, the maximum stress of 0.02 GPa, and the maximum atomic displacement of $5\times10^{-4}$ Å.

The CASTEP calculates the elastic properties from the first-principles using the finite strain theory [11], which gives the elastic constants as the proportionality coefficients relating the applied strain to the computed stress, $\sigma_i = C_{ij}\varepsilon_j$. From $C_{ij}$, the polycrystalline bulk modulus $B$ and shear modulus $G$ were further evaluated using the Voigt-Reuss-Hill approximation [12-14]. In addition, the Young's modulus $Y$, Poisson's ratio $v$, and shear anisotropy factor $A$ were estimated using the equations $Y = (9GB)/(3B + G)$, $v = (3B - 2G)/(6B + 2G)$ and $A = 4C_{44}/(C_{11} + C_{33} - 2C_{13})$, respectively.

In the present calculations, the spin-orbit coupling (SOC) has not been taken into consideration though the heavy transition metal Ir with 5d orbital is involved in this new compound. As far as the structural optimization, elastic properties and bonding characteristics are concerned, inclusion of SOC only has a minor effect. For example, the inclusion of the SOC in the calculations done for the transition metal based MAX phases, like $M_2AlC$ (M = Ti, V, and Cr) and $Mo_2AC$ (A = Al, Si, P, Ga, Ge, As, and In) has little or no influence on such properties [15, 16].

## 3. Results and discussion

### *3.1. Structural properties*

$Li_2IrSi_3$ crystallizes in the hexagonal structure with space group P6$_3$/mmc (194). The unit cell has 12 atoms containing two formula units (Fig. 1). The fully relaxed structure is obtained by optimizing the geometry with respect to lattice constants and internal atomic positions. The optimized Li atoms are situated in the 4*f* Wyckoff site with fractional coordinates (1/3, 2/3, 0.55889). The Si atoms are located on the 6*h* Wyckoff position with fractional coordinates (0.34291, 0.17145, 3/4). The Ir atoms are positioned in the 2*a* Wyckoff site with fractional coordinates (0, 0, 0). The calculated values of structural properties of $Li_2IrSi_3$ are presented in Table 1 along with the available experimental results. As can be seen from Table 1, the theoretical results are very close to the experimental values. This ensures the reliability of the present DFT-based first-principles calculations.

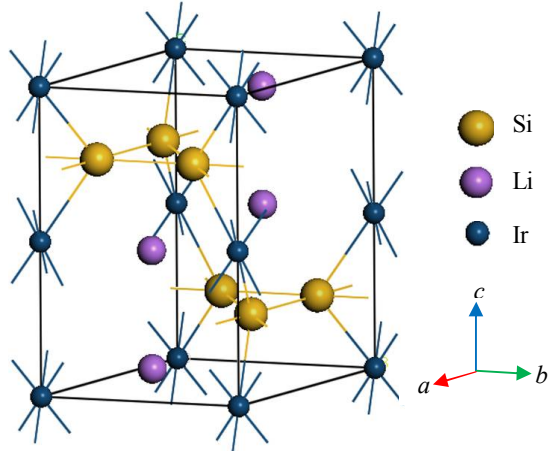

**Fig. 1.** The unit cell of hexagonal silicide superconductor $Li_2IrSi_3$.

**Table 1.** Lattice constants $a$ and $c$, internal parameter z, hexagonal ratio $c/a$, unit cell volume $V$, and bulk modulus $B$ with its pressure derivative $B'$ for $Li_2IrSi_3$.

| Properties | Expt. [2] | Present Calculation | Deviation (%) |
|---|---|---|---|
| $a$ (Å) | 5.01762 | 5.00690 | 0.21 |
| $c$ (Å) | 7.84022 | 7.86548 | 0.32 |
| Z | 0.56039 | 0.55889 | 0.27 |
| $c/a$ | 1.56254 | 1.57093 | 0.54 |
| $V$ (Å³) | 170.944 | 170.763 | 0.11 |
| $B$ (GPa) | -- | 115.899 | -- |
| $B'$ | -- | 4.24085 | -- |

### 3.2. *Elastic properties*

Table 2 lists our results for elastic properties at zero pressure. No experimental as well as theoretical data on elastic properties for $Li_2IrSi_3$ are available yet. Surprisingly, some of its elastic moduli match quite well with the results obtained for a MAX compound, $Ti_2CdC$ [17]. For comparison, these values are also listed in Table 2. Any mechanically stable hexagonal phase must conform to the conditions known as the Born criteria [18]: $C_{11} > 0$, $(C_{11} - C_{12}) > 0$, $C_{44} > 0$ and $(C_{11} + C_{12}) C_{33} > 2C_{13}^2$. The calculated elastic constants completely satisfy the aforementioned conditions. Thus, we can affirm that $Li_2IrSi_3$ is mechanically stable.

The small values of $C_{12}$ and $C_{13}$ imply that $Li_2IrSi_3$ should be brittle in nature [19]. The Pugh's criterion [20] and the Frantsevich's rule [21] also support this classification. According to Pugh's criterion, a material should be brittle if its Pugh's ratio $G/B > 0.5$, otherwise it should be ductile. The Frantsevich's rule has suggested the Poisson's ratio $v \sim 0.33$ as the critical value that separates the brittle and the ductile behavior. If the Poisson ratio $v$ is less than 0.33, the mechanical property of the material is dominated by brittleness, if larger than 0.33, the mechanical property of the material mainly shows ductility. From Table 2, it is evident that the compound $Li_2IrSi_3$ is brittle in nature like $Ti_2CdC$ that shows the general trend of MAX Phases [17, 22-24]. The relatively low value of the Poisson's ratio for $Li_2IrSi_3$ is indicative of its high degree of directional covalent bonding.

The difference between $C_{11}$ and $C_{33}$ suggests that $Li_2IrSi_3$ possesses anisotropy in elastic properties. Elastic anisotropy of a crystal reflects different characteristics of bonding in different directions. Essentially almost all the known crystals are elastically anisotropic, and a proper description of such an anisotropic behavior has, therefore, an important implication in engineering science and crystal physics since it correlates with the possibility of appearance of microcracks inside the crystals. The shear anisotropy factor for the {100} shear planes between the ⟨011⟩ and ⟨010⟩ directions is given by $A = 4C_{44}/(C_{11} + C_{33} - 2C_{13})$. For an isotropic crystal $A = 1$, while any value smaller or greater than unity is a measure of the degree of elastic anisotropy possessed by the crystal. The calculated shear anisotropic factor shown in Table 2 deviates slightly from unity, which implies that the in-plane and out-of-plane inter-atomic interactions differ only slightly. Another anisotropic factor is defined by the ratio between the linear compressibility coefficients along the *c*- and *a*-axis for the hexagonal crystal: $k_c/k_a = (C_{11} + C_{12} - 2C_{13})/(C_{33} - C_{13})$. This has also been calculated. The calculated value of 0.66 reveals that the compressibility along the *c*-axis is smaller than that along the *a*-axis for this new superconductor $Li_2IrSi_3$. The comparatively low shear modulus indicates the low coefficient of friction and good machinability of $Li_2IrSi_3$.

The ratio of the bulk modulus *B* to $C_{44}$ can be interpreted as a measure of plasticity [25]. Large values of $B/C_{44}$ indicate that the corresponding material have excellent lubricating properties. The plasticity may also be estimated by the values of $(C_{11} - C_{12})$ and Young's modulus *Y* [26]. The smaller values of $(C_{11} - C_{12})$ and *Y* indicate better plasticity of $Li_2IrSi_3$. The Debye temperature was $\Theta_D$ calculated from the elastic constants and was found to be 488 K. The $\Theta_D$ value estimated using the phonon contribution to the low temperature specific heat derived from experimental data on $C_P$ is 486 K [2]. Both the results are very close and it suggested that our calculations are reliable.

**Table 2**. Calculated single crystal elastic constants $C_{ij}$ (GPa), polycrystalline bulk modulus *B* (GPa), shear modulus *G* (GPa), Young modulus *Y* (GPa), Pugh's ratio *G/B*, Poisson's ratio *v*, and shear anisotropy factor *A* of $Li_2IrSi_3$.

| Single crystal elastic properties | | | Polycrystalline elastic properties | | |
|---|---|---|---|---|---|
| Properties | $Li_2IrSi_3$ | $Ti_2CdC$ | Properties | $Li_2IrSi_3$ | $Ti_2CdC$ |
| $C_{11}$ | 198 | 257 | B | 116 | 114 |
| $C_{12}$ | 71 | 68 | G | 72 | 64 |
| $C_{13}$ | 55 | 44 | Y | 179 | 162 |
| $C_{33}$ | 295 | 205 | G/B | 0.62 | 0.56 |
| $C_{44}$ | 68 | 36 | V | 0.24 | 0.26 |
| $C_{66}$ | 64 | 95 | A | 0.71 | 0.38 |

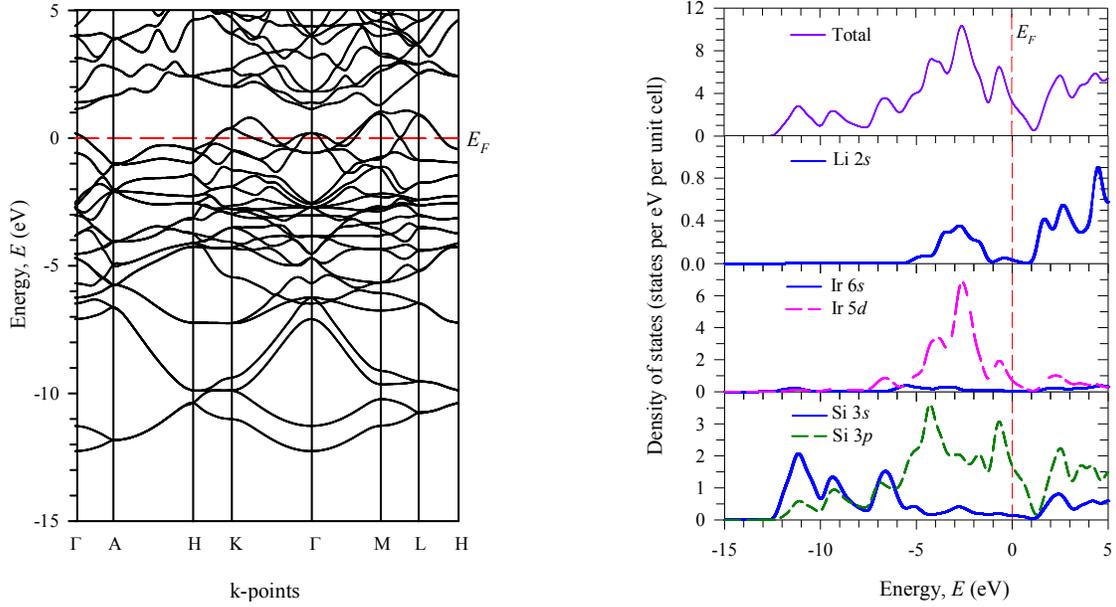

**Fig. 2.** Electronic structures of $Li_2IrSi_3$. (a) Band structure at equilibrium lattice parameters along the high symmetry wave-vectors, k and (b) total and partial energy densities of states. The Fermi level is set to 0 eV.

### 3.3. *Electronic properties*

The band structure of $Li_2IrSi_3$ with optimized lattice parameters along the high symmetry directions in the first Brillouin zone is presented in Fig. 2a. The Fermi level of this superconductor lies just below the valence band maximum near the $\Gamma$ point. The occupied valence bands spread widely from -12.3 eV to the Fermi level $E_F$. The valence band crosses the Fermi level making the system conducting. The highly dispersive nature of the band structure along the $\Gamma$-$A$ direction shows that electronic transport along *c*-direction is not that different from that in the plane. $Li_2IrSi_3$ should be treated as a three-dimensional, nearly isotropic system.

The total and partial energy density of states (DOS) of $Li_2IrSi_3$ is shown in Fig. 2b. The total DOS has a large value at the Fermi level $N(E_F)$, around 3.2 states per eV per unit cell, which is in reasonably good agreement with the value of 2.9 states per eV per unit cell found in literature [2]. It should be mentioned that, very recently Lu et al. [27] have reported electronic band structure calculations taking into SOC into consideration and has found $N(E_F)$ ~ 3.5 states per eV per unit cell. The DOS at the Fermi level arises mainly from Ir 5d and Si 3p states. Once again, this finding confirms the metallic conductivity of $Li_2IrSi_3$. The wide valence band consists of several distinct peaks. It is observed that the lowest-lying valence bands with two peak structures in the energy range between -12.3 and -7.5 eV are formed almost entirely by 3s and 3p states of Si. The next peak structure in the total DOS located between -7.5 and -5.8 eV is originated mainly from Si 3s, 3p and Ir 5d states. The second highest peak, between -5.8 and -3.3 eV, mainly consists of Si 3p and Ir 5d states. The highest peak in the range -3.3 to -1.3 eV is derived from the strong hybridization of Si 3s, 3p with Ir 5d states. Li 2s has little contribution. The intense peak just below the Fermi level is composed of the dominant contribution from Si 3p and Ir 5d states. It is evident that within the energy range from -12.3 eV to the Fermi level, a covalent interaction occurs between the constituting atoms due to the fact that states are degenerate with respect to both angular momentum and lattice site. It may be expected that the hybridization between 3s and 3p of Si in the lowest energy region should be responsible for the covalent Si-Si bond that form a three dimensional rigid network as a whole [2]. Further, a rather strong hybridization between Si 3p and Ir 5d states around the Fermi level would play the primary role in forming the covalent Ir-Si bond to crystallize the rigid structure of $Li_2IrSi_3$. Also, some ionic character can be expected due to the difference in the electronegativity among the comprising elements. Therefore, the bonding nature in $Li_2IrSi_3$ may be described as a combination of metallic, covalent and ionic.

## 3. Conclusions

We have calculated the structural parameters, elastic properties and electronic structures of newly discovered silicide superconductor $Li_2IrSi_3$ using DFT-based first-principles method. The structural properties obtained are in good agreement with the experimental values [2]. The calculated elastic constants allow us to conclude that the superconducting system $Li_2IrSi_3$ is mechanically stable. This new compound should have low coefficient of friction and good machinability due to its low shear modulus. In addition, $Li_2IrSi_3$ can be characterized as brittle material and shows a small elastic anisotropy. The electronic structures of $Li_2IrSi_3$ show that its bonding is a combination of covalent, ionic and metallic in nature. The rigid structure of $Li_2IrSi_3$ is the consequence of the strong covalent Ir-Si and Si-Si bonds. For covalent materials the electron-phonon coupling constant can be expressed as $\lambda = (M\langle\omega^2\rangle)^{-1}(dU/dz)^2 N(E_F)$ (the Hopfield expression), where $\omega$ is a characteristic phonon frequency and $dU/dz$ is the force due to electron-ion interaction. This equation, together with the weak coupling BCS equation for the superconducting transition temperature show that $T_c$ can be increased further by optimizing the values of $\omega$ and $N(E_F)$. The former has a direct link with the strength of the covalent bonds. Transition metal elements other than Ir might form stronger covalent bonds with Si 3p states and raise $\omega$ (~$\Theta_D$). $N(E_F)$, on the other hand, can be tuned by doping holes in the system which should result in a shift of the Fermi energy to a lower value and thereby increasing the EDOS at the Fermi level.